\documentclass[twoside]{article}
\usepackage{latexsym,amsmath,amssymb}
\numberwithin{equation}{section}

\newcommand{\zb}{\boldsymbol{b}}

\newcommand{\zh}{\boldsymbol{h}}

\newcommand{\zk}{\boldsymbol{k}}
\newcommand{\zl}{\boldsymbol{l}}

\newcommand{\zv}{\boldsymbol{v}}
\newcommand{\zx}{\boldsymbol{x}}

\newcommand{\zjm}{\boldsymbol{\jmath}}
\newcommand{\zell}{\boldsymbol{\ell}}

\newcommand{\zD}{\boldsymbol{D}}

\newcommand{\zQ}{\boldsymbol{Q}}
\newcommand{\zS}{\boldsymbol{S}}
\newcommand{\zT}{\boldsymbol{T}}
\newcommand{\zU}{\boldsymbol{U}}
\newcommand{\zV}{\boldsymbol{V}}

\newcommand{\zY}{\boldsymbol{Y}}
\newcommand{\zZ}{\boldsymbol{Z}}

\newfont{\gothic}{eufm10}% gothic letters
\newfont{\sgothic}{eufm7}%
\newcommand{\zcP}{\mathcal{P}}

\newcommand{\zcR}{\mathcal{R}}     % calligraphic letters

\newfont{\mmit} {cmmi10 scaled 1200}%
\newfont{\smit} {cmmi7 scaled 1200}%
\newfont{\ssmit}{cmmi5 scaled 1200}%

\newfont{\tenbfit}{cmmib10}%
\newfont{\svnbfit}{cmmib10 scaled 800}%
\newcommand{\zPi}{\mbox{\tenbfit\char'05\/}}%
\newcommand{\zSigma}{\mbox{\zPi}}%

\newfont{\tenbfsl}{cmbxti10}% <-- for idem tensor
\newcommand{\idem}{\mbox{\tenbfsl 1\/}}%

\newcommand{\bfnu}{\boldsymbol{\nu}}

\newcommand{\zcdot}{\cdot}   % bold center dot
\newcommand{\zdot}[1]{\skew1\dot{\mathnormal{#1}}}

\newcommand{\phat}[1]{{\skew3\hat{#1}}} % hat for slanted lettters
\newcommand{\xg}{\textsl{grad\:}}   % spatial grad
\newcommand{\xd}{\textsl{div\:}}% spatial div
\newcommand{\nuP}{\bfnu}
\newcommand{\nuQ}{\bfnu}

\newcommand{\tr}{\textsl{tr}\:}
\newcommand{\da}{\:\textsl{da}}
\newcommand{\dv}{\:\textsl{dv}}

\newcommand{\vis}{ {\!{\rm vis}} }

\newcommand{\kBT}{{k_{\rm B}\theta}}    % Boltzmann factor and temp.
\newcommand{\half}{\hbox{$\frac{1}{2}$}} % one half

\newcommand{\trans}{\scriptscriptstyle\top\mskip-4mu} % transpose

\newcommand{\conc}{c}

\begin{document}

\title{Microforces and the Theory of Solute Transport}
\author{\normalsize\sc Eliot
Fried{\raise1.25pt\hbox{$^{\scriptscriptstyle\dagger}$}}
\& Shaun Sellers{\raise1.25pt\hbox{$^{\scriptscriptstyle\ddagger}$}}
\\
\\
\small $^{\scriptscriptstyle\dagger}\mskip-6mu$ Department of Theoretical
and Applied
Mechanics
\\[-4pt] \small University of Illinois at Urbana-Champaign
\\[-4pt] \small Urbana, IL 61801-2935, USA
\\
\\[-12pt]
\small $^{\scriptscriptstyle\ddagger}\mskip-6mu$  School of Mathematics
\\[-4pt] \small University of East Anglia
\\[-4pt] \small Norwich NR4 7TJ, England}

\date{}%{ 21 March 1999}
\maketitle

\vspace{-250pt}
\begin{center}
\framebox{\footnotesize{\bf To appear in:} \emph{Zeitschrift f\"ur
angewandte Mathematik und Physik}}
\end{center}
\vspace{200pt}

\begin{abstract}\noindent
A generalized continuum framework for the theory of solute transport
in fluids is proposed and systematically developed. This framework rests
on the introduction of a generic force balance for the solute, a balance
distinct from the  macroscopic momentum balance associated with the
mixture. Special forms of such a force balance have been proposed and
used going back at least as far as Nernst's 1888 theory of diffusion.
Under certain circumstances, this force balance yields a Fickian
constitutive relation for the diffusive solute flux, and, in conjunction
with the solute mass balance, provides a generalized Smoluchowski
equation for the mass fraction.  Our format furnishes a systematic
procedure for generalizing convection-diffusion models of solute
transport, allowing for constitutive nonlinearities, external forces
acting on the diffusing constituents, and coupling between convection
and diffusion.
\end{abstract}

\thispagestyle{empty}

%%%%%%%%%%%%%%%%%%%%%%%%%%%%%
\section{Introduction}
\label{subsec:diffusion}

The transport of a solute in an incompressible fluid mixture
of \emph{density} $\varrho$ is commonly modeled by the
convection-diffusion equation\footnote{$\,$ Here, the operations of material
and spatial time-differentiation are denoted with a superposed dot and a
superposed grave accent, respectively; thus, for instance,
$\dot{\conc}=\grave{\conc}+\zv\!\zcdot\!\xg\conc$, with $\zv$ being the
\emph{mixture velocity}.}
\begin{equation}
\dot{\conc}
=\xd\bigl(D(\conc)\,\xg\conc\bigr),
\label{eq:diffusion}
\end{equation}
with $\conc$ the \emph{solute mass fraction} and $D\ge0$ the mass
fraction-dependent \emph{diffusivity}. This equation is  often derived
(see, for example, Landau \& Lifshitz \cite{Landau59}, de~Groot \& Mazur
\cite{Degroot83}, and Cussler \cite{Cussler84})  by adjoining to the
solute mass balance
\begin{equation}
\varrho\mskip1mu\dot{\conc}=-\xd\zjm,
\label{eq:mass}
\end{equation}
a constitutive relation
\begin{equation}
\zjm=-\varrho\mskip1mu D(\conc)\, \xg\conc
\label{eq:constitutive}
\end{equation}
of the sort proposed by Fick \cite{Fick55} for the \emph{diffusive
solute flux} $\zjm$.

Here, we propose and systematically develop an alternative approach to
solute transport. Our approach is based on the viewpoint  that to each
independent kinematic process there should correspond a system of
power-conjugate forces subject to a distinct force balance.\footnote{A
foundation for this viewpoint can be devised using the {\em principle of
virtual power}, see, for example, Germain \cite{Germain}, Antman \&
Osborn \cite{Antman}, Maugin \cite{Maugin}, and Fr\'emond \cite{Fremond}.
We follow here an alternative, but essentially equivalent, approach
taken by Ericksen \cite{E61}, Goodman \& Cowin \cite{GC72}, Capriz \&
Podio-Guidugli \cite{CP-G}, Capriz \cite{C89}, Fried \& Gurtin
\cite{FG}, Fried \cite{F}, and Gurtin \cite{G}.} Specifically, in the
context of solute transport, \emph{we identify mixture convection and
solute diffusion as independent kinematic processes}: mixture convection
is a macroscopic phenomenon described kinematically by the convective
mixture velocity $\zv$, whereas solute diffusion is a microscopic
phenomenon described kinematically at a continuum level  by the
diffusive solute flux $\zjm$, an object that measures the motion of
solute particles relative to the macroscopic motion of the mixture
comprised of solute and solvent. Further, to account for power
expenditures associated with the diffusive motion of atoms, we introduce
at a continuum level a system of forces conjugate to the diffusive
solute flux $\zjm$. Since they are associated with the microscopic
processes governing the diffusion of the solute, we use the expression
``microforce'' to identify these  forces. Additionally, we require that
these microforces be consistent with a \emph{solute microforce balance}.
This balance is a postulate ancillary to the conventional momentum
balance, which involves the forces that act power conjugate to the
convective mixture velocity $\zv$. Importantly, we do not assume that the
diffusive solute  flux $\zjm$ is determined by a constitutive
relation---Fickian or otherwise. Rather, \emph{we treat $\zjm$ as an
independent kinematic variable whose evolution is governed by the
microforce balance}.

There is ample historical precedence for basing solute transport on
additional, independent force balances. Nernst \cite{Nernst88},
following upon van't Hoff's \cite{Hoff87} analogy between the
\emph{osmotic pressure} $\varpi$ of the molecules of a sufficiently
dilute solute and  the pressure of an ideal gas, viewed diffusion as
driven by osmotic pressure gradients and developed a theory of diffusion
based on a balance of forces acting on the solute molecules.
Nernst proposed such force balances to
account for the influences of osmotic, viscous, electrostatic,
gravitational, centrifugal, and magnetic forces.  For example, for
electrolyte solutions, Nernst assumed separate  force balances for the
positive and negative ions.
Subsequently  Planck
\cite{Planck90} also exploited these ideas in his work on electrolyte
solutions. Further, in developing his theory of Brownian motion,
Einstein \cite{einstein05,einstein06,einstein08} employed force
balances on the Brownian particles. Neglecting the inertia of the
diffusing solute, we may write Nernst's solute force balance as
\begin{equation}
\zh+{\zell}=\frac{1}{n}\,\xg\varpi
           = \frac{M}{\varrho\mskip1mu\conc}\,\xg\varpi ,
\label{eq:nernst}
\end{equation}
wherein $\zh$ is the \emph{hydrodynamic drag force} acting on the solute
particles during motion through the solvent, $\zell$ represents the
\emph{external forces} acting on the solute particles,  $M$ is the
\emph{molecular weight} of the solute, and $n$ is the number
density---so that the right-hand side of (\ref{eq:nernst}) can be
interpreted as the \emph{osmotic force} per solute particle. Typically
the hydrodynamic drag force is taken to be linear in the solute velocity
relative to the mixture, whereby
\begin{equation}
\zh=-\zeta\mskip1mu\zjm,
\label{eq:frictional_force}
\end{equation}
with $\zeta$ a non-negative coefficient called the \emph{frictional
resistance} and $\zeta^{-1}$ called the \emph{mobility}.\footnote{Since
we take the solute flux $\zjm$ as a generalized velocity, our
definitions of frictional resistance and mobility differ from the usual
ones by a factor of $\varrho\conc$.} For an ideal solution, the \emph{osmotic
pressure} $\varpi$ is linear in the number density $n$, that is
\begin{equation}
\varpi=\kBT n=\frac{\varrho\mskip1mu\kBT\conc}{M},
\label{eq:osmotic}
\end{equation}
with $k_{\rm B}$ being \emph{Boltzmann's constant} and $\theta$ being the
\emph{absolute temperature}. Thus, in an isothermal, constant density
mixture,
\begin{equation}
\xg \varpi={\kBT}\,\xg{n}=\frac{\varrho\mskip1mu{\kBT}}{M}\xg\conc,
\label{eq:osmotic_grad}
\end{equation}
which, in conjunction with (\ref{eq:nernst}) leads to the flux
\begin{equation}
\zjm=-\frac{\kBT}{\conc\zeta}\,\xg\conc+\frac{\zell}{\zeta}.
\end{equation}
Hence, provided that external forces are absent, the balance
(\ref{eq:nernst}) and constitutive relations (\ref{eq:frictional_force})
and (\ref{eq:osmotic_grad}) yield an expression similar to the Fickian
constitutive equation (\ref{eq:constitutive}) with the classical relation
(see Nernst \cite[Page 615]{Nernst88})
\begin{equation}
D=\frac{\kBT}{\varrho\conc\mskip1mu\zeta},
\label{eq:diffusion_coef}
\end{equation}
between diffusivity and frictional resistance (or mobility).

There is, however, disagreement over the form and interpretation of the
solute force balance. For example, Einstein \cite[\S~4]{einstein06b}
asserted that the osmotic pressure is a ``virtual'' force, so that it
cannot be treated as a force acting on the individual
molecules.\footnote{The exact meaning of this statement is unclear to
the authors.} In fact, Einstein \cite{einstein06},  in his original
treatment of Brownian motion, proposed instead a fictitious force $\zk$
that acted on the individual particles but that balanced the osmotic
pressure gradient:
\begin{equation}\zk=\frac{M}{\varrho\mskip1mu\conc}\,\xg \varpi.
\end{equation}
Einstein \cite{einstein06} then took the solute force balance as
\begin{equation}\zh+\zl=\zk
\end{equation}
and thereby claimed (see Einstein \cite{einstein06b}) to have resolved
the apparent difficulty of having a virtual force act on the individual
particles.

Additionally, Debye \cite[Page~106]{Debye62} asserted that osmotic
forces are not ``true'' mechanical forces, so that the solute force
balance should be instead
\begin{equation}
\zh+{\zell}={\bf0}.
\label{eq:debye}
\end{equation}
In this alternative point of view, the flux $\zjm$ is decomposed as a sum
\begin{equation}
\zjm=\zjm_D+\zjm_F
\label{eq:debye_flux}
\end{equation}
involving a purely diffusive part $\zjm_D$  (usually called the
``ordinary diffusion'') and a part $\zjm_F$ due to external forces
(usually called the ``forced diffusion''). Commonly, Fick's first law
is invoked for the ordinary diffusive flux, so that
\begin{equation}
\zjm_D=-D\xg c,
\label{eq:debye_flux1.5}
\end{equation}
and the flux due to the external forces is assumed to have the form
\begin{equation}
\zjm_F=\frac{\zell}{\zeta},
\label{eq:debye_flux2}
\end{equation}
which follows from (\ref{eq:frictional_force}) and (\ref{eq:debye}).
Combining (\ref{eq:debye_flux2}) with the balance (\ref{eq:mass}) leads
to the \emph{generalized convection-diffusion equation}
\begin{equation}
\dot{\conc}=\xd \bigl(D\xg\conc-\frac{\zell}{\varrho \zeta}\bigr),
\label{eq:debye_mass}
\end{equation}
typically referred to as a Smoluchowski \cite{smoluchowski15} equation.
The diffusivity $D$ can be determined by considering the special case
where the external forces are given by a potential, viz.,
\begin{equation}
\zell=-\,\xg{V},
\label{eq:potential}
\end{equation}
and requiring the steady-state solution to correspond to that of a
Boltzmann \cite{boltzmann72} distribution, so that
\begin{equation}
\conc=\conc_0\exp\big(\frac{-V}{\kBT}\bigr).
\label{eq:boltzmann}
\end{equation}
This yields the relation (\ref{eq:diffusion_coef}), which is usually
enforced even when (\ref{eq:boltzmann}) no longer holds.

This latter format was used by Debye \cite{debye13,debye29} in his
theory of rotary diffusion in  dielectrics to include the effect of the
external electric field on the  diffusing dipoles. It is still commonly
used to obtain extensions of the Fickian constitutive relation to
include, for instance, the effects of external electric fields on
electrolytes. Additionally it is used to model the interaction of
diffusion and external fields on dilute suspensions (see, for example,
Giesekus \cite{Giesekus58} and, more recently, Doi \& Edwards
\cite{Doi86}). However, Bird, Armstrong, \&  Hassager \cite{Bird77} based
their approach to polymeric suspensions on a force balance such as
(\ref{eq:nernst}) that explicitly includes a contribution attributed to
the Brownian forces on the polymer molecule (see, for example,
\cite[Equation~13.2-1]{Bird77}).

Interestingly, in the case of linear constitutive relations, the
approaches described above lead to the same resulting equations---namely
(\ref{eq:debye_mass}) and (\ref{eq:diffusion_coef}). However, the
assumptions involved are rather different. In particular, it is not clear
how to generalize either approach to allow for constitutive
nonlinearities. For example, in (\ref{eq:debye_flux}) the ordinary
diffusive flux $\zjm_D$ is determined by a constitutive relation,
whereas the forced diffusive part $\zjm_F$ is determined by a balance
law---a procedure highly atypical for a continuum theory.  Furthermore,
the validity of (\ref{eq:diffusion_coef}), derived from an assumed
steady-state solution, is not clear in unsteady situations.

In our opinion, these issues can be resolved by a more general treatment
of  the notion of forces in a binary mixture. Consequently, in the
following, we develop a theoretical framework for diffusion  based on a
\emph{generic} microforce balance without a priori assumptions on the
nature of the microforces or of steady-state concentration profiles.
Furthermore, we allow for constitutive nonlinearities and external
fields. We follow the procedure of modern thermodynamics, wherein
consistency with the second law restricts the form of the constitutive
relations. Our proposed format provides a systematic procedure for the
generalization of convection-diffusion equation models for solute
transport. In particular, it leads to a force balance that  explicitly
includes the diffusive or Brownian forces acting on the solute. Only in
special cases, such as linear constitutive relations, can this force
balance be expressed in the oft used form (\ref{eq:debye}) with the
corresponding decomposition  (\ref{eq:debye_flux}) of the flux proposed
by Debye.

For clarity, we restrict our attention to purely mechanical processes
in a binary mixture of solute and solvent, ignoring thermal and other
effects. In this context, the first and second laws of thermodynamics
are replaced by a \emph{free-energy imbalance} that accounts for power
expenditures associated not only with mixture convection but also with
solute diffusion. That diffusion does, indeed, lead to the expenditure of
power was clearly recognized and utilized by van't Hoff \cite{Hoff87}
as an experimental means of measuring the osmotic pressure. First, we
consider simple diffusion without mixture convection. This special case
allows us to develop the theory without the complications of coupling
with the macroscopic mixture motion. We then treat both convection and
diffusion.

%%%%%%%%%%%%%%%%%%%%%%%%%%%%%%%%%%%%%%%%%%%%%%%%%%%%%%%%
%%%%%%%%%%%%%%%%%%%%%%%%
\section{Simple theory accounting only for diffusion}

We consider a binary mixture of solute and solvent that occupies a fixed region
$\zcR$ of three-dimensional space. In the present context, where we ignore
the macroscopic motion of the mixture, the mixture density is constant.
Without loss of generality, we take that constant to be unity.

\subsection{Basic laws}
\label{sect:balance}

The simple theory is based upon the following basic laws:
\begin{itemize}
\item[$\bullet$] solute mass balance;
\item[$\bullet$] solute microforce balance;
\item[$\bullet$] mixture free-energy imbalance.
\end{itemize}
We formulate these laws in global form for an arbitrary subregion $\zcP$ of
$\zcR$. The outward unit-normal field on the boundary $\partial\zcP$ of
$\zcP$ is denoted by $\nuQ$.

%%%%%%%%%%%%%%%%%%%%%%%%%%%%%
\subsubsection{Solute mass balance}
\label{subsect:solute_mass}

We introduce fields\footnote{Since the mixture density is assumed to be unity,
there is no distinction between solute mass and solute mass fraction, or
between solute mass flux and solute mass fraction flux.}
\begin{center}
\begin{tabular}{cl}
$\conc$    & \emph{solute mass fraction},                   \\ [0.5ex]
$\zjm$     & \emph{diffusive solute  flux},     \\ [0.5ex]
$m$        & \emph{external solute supply},     \\ [0.5ex]
\end{tabular}
\end{center}
in which case the integrals
\begin{equation}
\int\limits_{\zcP}%\varrho\mskip1mu
                  \conc\dv,
\qquad\qquad
\int\limits_{\partial\zcP}\zjm\!\cdot\!\nuQ\da,
\qquad\text{and}\qquad
\int\limits_{\zcP}m\dv
\end{equation}
represent, respectively, the solute mass in $\zcP$,
the solute mass added to
$\zcP$, per unit time, by diffusion across $\partial\zcP$, and the solute
mass added to $\zcP$, per unit time,
by external agencies.

\emph{Solute mass balance} is the postulate that, for each subregion
$\zcP$ of $\zcR$ and each instant,
\begin{equation}
\grave{\overline{\int\limits_{\zcP}%\varrho\mskip1mu
                                    \conc\dv}}
=-\int\limits_{\partial\zcP}\zjm\!\cdot\!\nuQ\da
+\int\limits_{\zcP}m\dv.
\label{eq:solute_mass_balance}
\end{equation}
Equivalently, since $\zcP$ is arbitrary, we may enforce solute mass balance
via the local field equation
\begin{equation}
%\varrho\mskip1mu
\grave{\conc}=-\xd\zjm+m.
\label{eq:local_solute_mass_balance}
\end{equation}

%%%%%%%%%%%%%%%%%%%%%%%%%%%%%%%%%%%%%%%%
\subsubsection{Solute microforce balance}
\label{subsect:micromoment}

Associated with the evolution of the diffusive solute flux $\zjm$, we
introduce a system of power-conjugate \emph{microforces}, consisting of
\begin{center}
\begin{tabular}{cl}
$\zSigma$ & \emph{solute microstress tensor},       \\[0.5ex]
$\zh$     & \emph{internal solute body microforce}, \\[0.5ex]
$\zell$   & \emph{external solute body microforce}, \\[0.5ex]
\end{tabular}
\end{center}
in which case the integrals
\begin{equation}
\int\limits_{\zcP}\zSigma\nuQ\da,
\qquad\qquad
\int\limits_{\zcP} \zh\dv,
\qquad\text{and}\qquad
\int\limits_{\zcP}\zell\dv
\end{equation}
represent, respectively, the solute microforces exerted on the region $\zcP$
by the solute microtraction distributed over $\partial\zcP$, by agencies
within $\zcP$, and by agencies external to $\zcP$, respectively.  In the
present mixture context, internal forces arise from  the action of  the
solvent on the solute, or from solute-solute interactions. External forces
are those arising from sources remote from the mixture, such as those
arising from applied electromagnetic or gravitational fields. The
microstress tensor $\zSigma$ allows for surface tractions that are not
parallel to the surface normal.
We assume that the inertia associated with solute diffusion is negligible
and, hence, impose a \emph{solute microforce balance}, which is the
postulate that, for each subregion $\zcP$ of $\zcR$ and each instant,
the resultant microforce acting on $\zcP$ must vanish:
\begin{equation}
\int\limits_{\partial\zcP}\zSigma\nuQ\da
+\int\limits_{\zcP}(\zh+\zell)\dv={\bf0}.
\label{eq:orientational_balance}
\end{equation}
Equivalently, since $\zcP$ is arbitrary, we may enforce solute microforce
balance via the local field equation
\begin{equation}
\xd\zSigma+\zh+\zell={\bf0}.
\label{eq:local_microforce_balance}
\end{equation}
 A comparison to the Nernst force balance
 (\ref{eq:nernst}) suggests an interpretation of the
quantity $\zSigma$ as a generalized stress tensor that allows for the
possibility of
tangential osmotic forces on a semipermeable membrane.
At this point, we
impose no constitutive assumptions concerning the microforces. In particular,
we allow for the possibility that $\zSigma$ vanish identically, which would
correspond to the alternative balance (\ref{eq:debye}) proposed by Debye.

%%%%%%%%%%%%%%%%%%%%%%%%%%%%%%%%%%%%%%%%%%%%%%%%%%%%%%%%%%%%%%%%%%%%%%%%%%%%%
\subsubsection{Free-energy imbalance}
\label{subsect:energy}

In this purely mechanical theory, the first and second laws of
thermodynamics are replaced by a \emph{free-energy imbalance}. To
formulate this imbalance, we introduce fields
\begin{center}
\begin{tabular}{cl}
$\psi$ & \emph{mixture free-energy density},     \\[0.5ex]
$\mu$  & \emph{solute diffusion potential}, \\[0.5ex]
\end{tabular}
\end{center}
in which case that the integrals
\begin{equation}
\int\limits_{\zcP}%\varrho\mskip1mu
                   \psi\dv
\qquad\text{and}\qquad
\int\limits_{\zcP}\mu\mskip1mu{m}\dv
\end{equation}
represent, respectively, the free-energy in $\zcP$ and the rate at which
free-energy is added to $\zcP$ by the external supply of solute to $\zcP$.

Further, the integrals
\begin{equation}
\int\limits_{\partial\zcP}\zSigma\nuP\!\zcdot\!\zjm\da
\qquad\text{and}\qquad
\int\limits_{\zcP}\zell\!\zcdot\!\zjm\dv
\end{equation}
provide an accounting of the power expended on $\zcP$ by the solute
microtraction distributed over $\partial\zcP$ and by the agencies acting
external to that region. These expressions represent the power expended by
the diffusion of the solute particles.

Consistent with our assumption that solute inertia is negligible, we
neglect the kinetic energy of solute diffusion. \emph{Free-energy
imbalance} is, then, the postulate that, for each subregion $\zcP$ of $\zcR$
and each instant:
\begin{equation}
{\grave{\overline{\int\limits_{\zcP}%\varrho\mskip1mu
                                    \psi\dv}}}
\ \le\
\int\limits_{\partial\zcP}\zSigma\nuP\!\zcdot\!\zjm\da
+\int\limits_{\zcP}(\zell\!\zcdot\!\zjm+\mu\mskip1mu{m})\dv.
\label{eq:difference}
\end{equation}
Equivalently, since $\zcP$ is arbitrary, we may enforce free-energy
imbalance via the local field inequality
\begin{equation}
-%\varrho\mskip1mu
\grave{\psi}+%\varrho\mskip1mu
               \mu\mskip1mu\grave{\conc}
-\zh\!\zcdot\!\zjm+(\zSigma+\mu\idem)\!\zcdot\!\xg\zjm\ge0.
\label{eq:local_strong_energy_imbalance_2}
\end{equation}

%%%%%%%%%%%%%%%%%%%%%%%%%%%%%%%%%%%%%%%
\subsection{Constitutive equations}
\label{sect:const}

The preceding balance laws do not form a determinate system of equations.
We now develop constitutive equations that will provide closure relations.
We assume that the free-energy density $\psi$, the solute diffusion
potential $\mu$, the solute microstress tensor $\zSigma$, and the internal
solute body microforce $\zh$ are determined constitutively as functions of the
solute mass fraction $c$ and the diffusive solute flux $\zjm$, so that
\begin{equation}
(\psi,\mskip2mu\mu,\mskip2mu\zSigma,\mskip2mu\zh)
=\bigl(\hat{\psi}(\conc,\zjm),\mskip2mu\hat{\mu}(\conc,\zjm),
\mskip2mu\phat{\zSigma}(\conc,\zjm),\mskip2mu\hat{\zh}(\conc,\zjm)\bigr).
%\left.
%\begin{split}
%\psi&=\hat{\psi}(\conc,\zjm), \qquad
%\zSigma=\hat{\zSigma}(\conc,\zjm),\\
%\mu&=\hat{\mu}(\conc,\zjm),
% \qquad  \mskip9mu
%\zh=\hat{\zh}(\conc,\zjm).
%\end{split} \quad
%\right\}
\label{eq:cr}
\end{equation}
We emphasize that, in contrast to the standard approach to the theory of
mass diffusion, the diffusive flux $\zjm$ is not given by a
constitutive relation.

Inserting the constitutive relations (\ref{eq:cr}) into the local
free-energy imbalance (\ref{eq:local_strong_energy_imbalance_2}), we arrive
at the functional inequality
\begin{multline}
%\varrho
\bigl(\frac{\partial\hat{\psi}}{\partial\conc}(\conc,\zjm)
-\hat{\mu}(\conc,\zjm)\bigr)\grave{\conc}
+%\varrho
\frac{\partial\hat{\psi}}{\partial\zjm}
(\conc,\zjm)\!\zcdot\!\skew3\grave{\zjm}
%%+\frac{\partial\hat{\psi}}{\partial\xg \conc}(z)\!\zcdot\!
%%\zdot{\overline{\xg \conc}}
+\hat{\zh}(\conc,\zjm)\!\zcdot\!\zjm
%\\[4pt]
-\bigl(\phat{\zSigma}(\conc,\zjm)+\hat{\mu}(\conc,\zjm)\idem\bigr)
\!\zcdot\!\xg\zjm
\le0,
\label{eq:dissipation_imbalance_with_constitutive_relations}
\end{multline}
and, following the procedure founded by Coleman \& Noll \cite{CN} in their
incorporation of the second law into continuum thermomechanics, conclude
that:
\begin{itemize}
\item[({\em i\/})] the constitutive response functions $\hat{\psi}$,
$\phat{\mu}$, and $\phat{\zSigma}$ delivering the free-energy density
$\psi$, the diffusion potential $\mu$, and the stress $\zSigma$ must be
independent of the diffusive mass flux $\zjm$, with\footnote{Here, a
prime indicates with respect to the argument of a function depending on
a scalar field. It is noteworthy that the solute microstress reduces to
an isotropic tensor because we do not include the gradient $\xg\zjm$ of
the diffusive solute flux $\zjm$ in the list of independent constitutive
variables.}
\begin{equation}
\psi=\hat{\psi}(c),
\qquad
\mu=\hat{\psi}{}^{\prime}(\conc),
\qquad\text{and}\qquad
\zSigma=-\hat{\psi}{}^{\prime}(\conc)\idem;
\label{eq:basic_model_restrictions}
\end{equation}
\item[({\em ii\/})] the constitutive response function $\hat{\zh}$ for the
internal solute body microforce $\zh$ must be consistent with the
\emph{residual inequality}
\begin{equation}
\hat{\zh}(\conc,\zjm)\!\zcdot\!\zjm\le0.
\label{eq:residual_inequality}
\end{equation}
\end{itemize}

Granted smoothness of the response function $\hat{\zh}$, a result
appearing in Gurtin \& Voorhees \cite{gurtin93} yields the general
solution
\begin{equation}
\hat{\zh}(\conc,\zjm)=-\zZ(\conc,\zjm)\mskip1mu\zjm
%=-\boldsymbol{\sf Z}(\conc,\zjm)\mskip1mu\zjm
\label{eq:solution_of_residual_inequality}
\end{equation}
of (\ref{eq:residual_inequality}), where the symmetric \emph{generalized
frictional resistance tensor} $\zZ$ must obey
\begin{equation}
\zjm\!\zcdot\!\zZ(\conc,\zjm)\mskip1mu\zjm\ge0.
%\zjm\!\zcdot\!\boldsymbol{\sf Z}(\conc,\zjm)\,\zjm\ge0.
\end{equation}
Objectivity further requires that under the transformation
\begin{equation}
c\mapsto c,
\qquad
\zjm\mapsto\zQ\zjm,
\end{equation}
with $\zQ$ an arbitrary rotation, the quantities $\psi$, $\mu$, $\zSigma$,
$\zh$, and $\zell$ transform as
\begin{equation}
\psi\mapsto\psi,
\qquad
\mu\mapsto\mu,
\qquad
\zSigma\mapsto\zQ\zSigma\zQ^{\trans},
\qquad
\zh\mapsto\zQ\zh,
\qquad\zell\mapsto\zQ\zell,
\end{equation}
which, in particular, implies that
\begin{equation}
\zZ(\conc,\zjm)=\zeta(\conc,\jmath)\idem,
\qquad\jmath=|\zjm|,
%\boldsymbol{\sf Z}(\conc,\zjm)=\zeta(\conc,\zjm)\mskip1mu\idem
\label{eq:Z=zeta_idem}
\end{equation}
where the \emph{scalar frictional resistance} $\zeta$ must obey
\begin{equation}
\zeta(\conc,\jmath)\ge0
\end{equation}
for all $(\conc,\jmath)$. Combining (\ref{eq:solution_of_residual_inequality})
and (\ref{eq:Z=zeta_idem}), we see that
\begin{equation}
\zh=-\zeta(c,\jmath)\mskip1mu\zjm,
\end{equation}
so that the internal solute body microforce $\zh$ must be collinear with the
diffusive solute flux $\zjm$.

The foregoing results show that the  behavior of a medium of the sort
considered here is completely determined by the provision of two constitutive
response functions: (\emph{i}) $\hat{\psi}$ determining the mixture
free-energy density as a function of the solute mass fraction $\conc$;
and, (\emph{ii}) the scalar frictional resistance $\zeta$, which
determines the internal solute body microforce $\zh$ and, in general, may
depend on both the solute mass fraction $\conc$ and the magnitude
$\jmath$ of the diffusive solute flux $\zjm$.

The governing equations that arise on substituting the foregoing
thermodynamically consistent constitutive equations in the local field
equations (\ref{eq:local_solute_mass_balance}) and
(\ref{eq:local_microforce_balance}) expressing solute mass balance and solute
microforce balance are
\begin{equation}
\left\delimiter0
\begin{split}
%\varrho\mskip1mu
       \grave{\conc}+\xd\zjm&=m,
\\[4pt]
\zeta(\conc,\jmath)\mskip1mu\zjm
+\xg\bigl(\hat{\psi}{}^{\prime}(\conc)\bigr)&=\zell.
\quad
\end{split}
\right\}
\label{eq:basic_model_final_equations}
\end{equation}

Provided that $\zeta$ is nonvanishing,
(\ref{eq:basic_model_final_equations})$_2$ yields an expression
\begin{equation}
\zjm=
-\frac{\hat{\psi}{}^{\prime\prime}(\conc)}{\zeta(\conc,\jmath)}\,
\xg\conc
+\frac{\zell}{\zeta(\conc,\jmath)},
\label{eq:expression_for_flux}
\end{equation}
for the diffusive solute flux $\zjm$. This leads, in conjunction with
(\ref{eq:basic_model_final_equations})$_1$, to the \emph{generalized
Smoluchowski equation}
\begin{equation}
%\varrho\mskip1mu
         \grave{\conc}=\xd\big(
\frac{\hat{\psi}{}^{\prime\prime}(\conc)}{\zeta(\conc,\jmath)}\,\xg\conc
-\frac{\zell}{\zeta(\conc,\jmath)}\big).
\label{eq:smoluchowski}
\end{equation}
Note that, when  the frictional resistance $\zeta$ depends on
$\jmath$, the evolution equation (\ref{eq:smoluchowski}) \emph{does not}
decouple from the microforce balance
(\ref{eq:basic_model_final_equations})$_2$. In this case, the flux
$\zjm$ does not separate into distinct ordinary and forced  diffusive
contributions. Comparing (\ref{eq:expression_for_flux}) with the
conventional relation (\ref{eq:constitutive}), we obtain a generalized
expression
\begin{equation}
D(c,\jmath)=\frac{\hat{\psi}{}^{\prime\prime}(c)}{\zeta(c,\jmath)}
\label{eq:generalized_stokes_einstein}
\end{equation}
for the diffusivity, which is not predicated on a particular
steady-state solution and, moreover, is not restricted to the linear
regime. Hence, within our framework, the Fickian constitutive
relation (\ref{eq:constitutive}) can be interpreted as an expression of the
solute microforce balance, granted that the internal body microforce is
linear in the diffusive solute flux and that the external solute body
microforce vanishes. In the special case governed by (\ref{eq:potential}),
the steady-state solution is
\begin{equation}
\mu=\hat{\psi}^\prime(c)=-V+\text{constant},
\label{eq:solution}
\end{equation}
which need not correspond to the Boltzmann distribution
(\ref{eq:boltzmann}).

Our results for the microforce balance are consistent with the
Nernst balance (\ref{eq:nernst}). In fact, a comparison of
(\ref{eq:nernst}) to (\ref{eq:basic_model_final_equations})$_2$ shows
that they are identical in form if
\begin{equation}
\varpi^\prime(c)=\frac{\conc}{M}\hat{\psi}^{\prime\prime}(c)
=\frac{\conc}{M}\hat{\mu}^\prime(c).
\label{eq:osmotic_eq}
\end{equation}
Thus,
(\ref{eq:basic_model_restrictions}) shows that the osmotic traction acts only
normal to a semipermeable membrane and is determined by the mixture
free-energy density. Note that (\ref{eq:basic_model_final_equations})$_2$
is consistent with the Debye (\ref{eq:debye}) balance only when the
frictional resistance $\zeta$ is independent of the flux $\zjm$.

%%%%%%%%%%%%%%%%%%%%%%%%%%%%%%%%%%%%%%%%%%%%%%%%%%
\subsection{Special case: diffusion of non-interacting spheres}
\label{subsubsect:ideal}

We now specialize the preceding results to the case of diffusing,
non-interacting spherical particles, which corresponds to the dilute Brownian
spheres originally treated by Einstein \cite{einstein05}. Our framework
requires only the specification of constitutive relations for the mixture
free-energy density and the frictional resistance. We stipulate that
$\psi$ be determined by a response function $\hat{\psi}$ of the form
\begin{equation}
\hat{\psi}(\conc)=\kBT\mskip1mu%\varrho\mskip1mu
                               \conc\log{\conc},
\label{eq:assumption1b}
\end{equation}
which corresponds to an \emph{ideal mixture} in the sense that it
consists (to within an arbitrary additive constant) only of a classical
\emph{entropic} contribution, a contribution that drives diffusion. This
choice does not encompass particle-particle interactions.  Insertion of
(\ref{eq:assumption1b}) into (\ref{eq:osmotic_eq}) leads to
\begin{equation} \varpi(c)=\frac{\kBT \mskip1mu \conc}{M}.
\end{equation}
Further,
we require that the response function $\hat{\zh}$ be linear in the
diffusive solute flux $\zjm$ with the specific form
\begin{equation}
\zh=\hat{\zh}(c,\zjm)=-\zeta(\conc)\mskip1mu\zjm,
\qquad\text{with}\qquad\zeta(\conc)=
  \frac{6\pi\eta_s r}{%\varrho\mskip1mu
               \conc      },
\label{eq:assumption2b}
\end{equation}
so that the internal body microforce is consistent with Stokes'
\cite{Stokes} law for the drag on a rigid spherical particle slowly
translating in an incompressible Newtonian fluid. Here, $\eta_s$ is the
shear viscosity of the pure fluid solvent and $r$ is the particle
radius. Finally, we suppose that the external body microforce is
generated by the gradient of a potential according to
(\ref{eq:potential}).

Granted these assumptions, we may use (\ref{eq:expression_for_flux}) to
obtain
\begin{equation}
\zjm=
-\frac{\kBT{%\varrho\mskip1mu
             }}{6\pi\eta_s r}\,\xg \conc
-\frac{%\varrho\mskip1mu
                        c  }{6\pi\eta_s r}\,\xg V,
\label{eq:debye_eq}
\end{equation}
which, when inserted into the mass balance
(\ref{eq:basic_model_final_equations})$_2$ yields the generalized
diffusion equation
\begin{align}
%\varrho\mskip1mu
                  \grave{\conc}
%&=D\mskip1mu\xd\bigl(\conc\,\xg(\log{\conc}+\frac{V}{\kBT})\bigr)
%\notag\\[4pt]
&=D\mskip1mu\xd\bigl(\xg\conc+\frac{c}{\kBT}\,\xg{V}\bigr),
\label{eq:debye_eq1}
\end{align}
with the diffusivity $D$ having the explicit form
\begin{equation}
D=\frac{%\varrho
         \kBT}{6\pi\eta_s r}
\label{eq:stokes-einstein}
\end{equation}
consistent with the diffusivity expression originally derived by Einstein
\cite{einstein05} for Brownian spheres. Thus, the contribution of $\psi$ in
(\ref{eq:debye_eq}) leads to a classical, linear second-order diffusion
term that tends to randomize concentrations, which justifies our description
of it as \emph{entropic}. Additionally the steady-state mass fraction has
the classical Boltzmann form (\ref{eq:boltzmann}).

Importantly, our derivation of (\ref{eq:debye_eq1}) and
(\ref{eq:stokes-einstein}) relies neither on a Fickian constitutive relation
for the diffusive mass flux nor on an assumed form for the mass fraction
distribution in equilibrium. Rather, the relation (\ref{eq:stokes-einstein})
is a \emph{consequence} of the thermodynamic theory and the specialized
ideal constitutive expressions (\ref{eq:assumption1b}) and
(\ref{eq:assumption2b}).

%%%%%%%%%%%%%%%%%%%%%%%%%%%%%%%%%%%%%%
\section{Theory accounting for both mass diffusion and convection}
\label{subsect:generalization}

We now account for motion of the mixture as described by a convective
velocity $\zv$. Further, we allow the region $\zcR$ of three-dimensional
space occupied by the mixture to evolve with time.

%%%%%%%%%%%%%%%%%%%%%%%%%%%%%%%%%%%%%%%%%%%%%
\subsection{Basic laws}
\label{sect:balance2}

The theory is based upon the following laws:
\begin{itemize}
\item[$\bullet$] mixture mass balance;
\item[$\bullet$] solute mass balance;
\item[$\bullet$] mixture momentum balance;
\item[$\bullet$] moment of mixture-momentum balance;
\item[$\bullet$] solute microforce balance;
\item[$\bullet$] mixture free-energy imbalance.
\end{itemize}
As in our treatment of the theory without mixture convection, we formulate
the basic laws in global form for an arbitrary \emph{material} subregion
$\zcP$ of $\zcR$.

%%%%%%%%%%%%%%%%%%%%%%%%%%%%%%%%%%%
\subsubsection{Mixture mass balance}
\label{subsect:mixture_mass}

We now allow for convection and  variations in the mixture mass density,
so that we introduce fields
\begin{center}
\begin{tabular}{cl}
$\varrho$ & \emph{mixture mass density}, \\ [0.5ex]
$\zv$   & \emph{mixture velocity}.        \\ [0.5ex]
\end{tabular}
\end{center}

\emph{Mixture mass balance} is the postulate that, for each material
subregion $\zcP$ of $\zcR$ and each instant,
\begin{equation}
\grave{\overline{\int\limits_{\zcP}\varrho\dv}}=0.
\label{eq:mixture_mass_balance}
\end{equation}
Equivalently, since $\zcP$ is arbitrary, we may enforce mixture mass
balance via the local field equation
\begin{equation}
\dot{\varrho}+\varrho\,\xd\zv=0.
\label{eq:local_mixture_mass_balance}
\end{equation}

%%%%%%%%%%%%%%%%%%%%%%%%%%%%%
\subsubsection{Solute mass balance}
\label{subsect:solute_mass2}

With mixture convection taken into account, the global statement of
solute mass balance for a material subregion $\zcP$ of $\zcR$ has the
form (\ref{eq:solute_mass_balance}) considered in the absence of
convection, provided that $\conc$ is replaced by $\varrho\mskip1mu\conc$.
Thus, bearing in mind the mixture mass balance
(\ref{eq:local_mixture_mass_balance}), the local equivalent of solute
mass balance now reads
\begin{equation}
\varrho\mskip1mu\zdot{\conc}=-\xd\zjm+m.
\label{eq:local_solute_mass_balance2}
\end{equation}

%%%%%%%%%%%%%%%%%%%%%%%%%%%%%%%%%%%%%%%%
\subsubsection{Solute microforce balance}
\label{subsect:micromoment2}

With mixture convection taken into account, the global and local
statements, (\ref{eq:orientational_balance}) and
(\ref{eq:local_microforce_balance}), of solute microforce balance remain
unchanged from those introduced earlier.

%%%%%%%%%%%%%%%%%%%%%%%&%%%%%%%%%%%%%%
\subsubsection{Mixture momentum balance and moment of mixture-momentum
balance}
\label{sect:momentum}

Associated with the convective mixture velocity, we introduce fields
\begin{center}
\begin{tabular}{cl}
$\zT$  &  \emph{mixture stress},              \\[0.5ex]
$\zb$  &  \emph{external mixture body force density}, \\[0.5ex]
\end{tabular}
\end{center}
in which case the integrals
\begin{equation}
\int\limits_{\zcP}\zT\nuQ\da
\qquad\text{and}\qquad
\int\limits_{\zcP}\zb\dv
\end{equation}
represent the mixture forces exerted on the region $\zcP$ by the mixture
traction distributed over $\partial\zcP$, by agencies within $\zcP$, and by
agencies external to $\zcP$, respectively.

In addition to the mixture mass balance
(\ref{eq:local_mixture_mass_balance}), the solute mass balance
(\ref{eq:local_solute_mass_balance2}), and the solute microforce balance
(\ref{eq:local_microforce_balance}), we enforce balances of mixture momentum
and momentum of mixture momentum.

\emph{Mixture momentum balance} is the postulate that, for each material
subregion $\zcP$ of $\zcR$ and each instant, the time-rate at which the
convective momentum within $\zcP$ changes be equal to the resultant of the
convective forces acting on $\zcP$:
\begin{equation}
\grave{\overline{\int\limits_{\zcP}\varrho\mskip1mu\zv\dv}}
=\int\limits_{\partial\zcP}\zT\nuQ\da
+\int\limits_{\zcP}\zb\dv.
\label{eq:cauchy_balance}
\end{equation}
Further, \emph{moment of mixture momentum balance} is the postulate that,
for each material subregion $\zcP$ of $\zcR$ and each instant, the time-rate
at which the moment of the convective momentum within $\zcP$ changes be equal
to the resultant of the convective torques acting on $\zcP$:
\begin{equation}
\grave{\overline{\int\limits_{\zcP}\zx\!\wedge\!\varrho\mskip1mu\zv\dv}}
=\int\limits_{\partial\zcP}\zx\!\wedge\!\zT\nuQ\da
+\int\limits_{\zcP}\zx\!\wedge\!\zb\dv.
\label{eq:moment_balance}
\end{equation}
Equivalent to these balances, we have the local field equations
\begin{equation}
\varrho\mskip1mu\skew2\dot{\zv}=\xd\zT+\zb
\qquad\text{and}\qquad
\zT=\zT^{\trans}.
\label{eq:local_cauchy_balance}
\end{equation}

%%%%%%%%%%%%%%%%%%%%%%%&%%%%%%%%%%%%%%
\subsubsection{Free-energy imbalance}
\label{sect:energy}

Our previous statement of free-energy imbalance must be modified to account
for variations in the mixture mass density, the kinetic energy of the
mixture, and the power expended by the mixture stress and external mixture
body force. Specifically, we now require that for each material subregion
$\zcP$ of $\zcR$ and each instant:
\begin{equation}
{\grave{\overline{\int\limits_{\zcP}\varrho(\psi+\half|\zv|^2)\dv}}}
\ \le\
\int\limits_{\partial\zcP}
\bigl(\zSigma^{\trans}\zjm+\zT^{\trans}\zv\bigr)\!\zcdot\!\nuP\da
+\int\limits_{\zcP}(\zell\!\zcdot\!\zjm+\mu\mskip1mu{m}+\zb\!\zcdot\!\zv)\dv.
\label{eq:difference2}
\end{equation}
Equivalently, since $\zcP$ is arbitrary, we may enforce mixture energy
imbalance via the local inequality
\begin{equation}
-\varrho\mskip1mu\dot{\psi}+\varrho\mskip1mu\mu\mskip1mu\dot{\conc}
-\zh\!\zcdot\!\zjm+\zT\!\zcdot\!\zD+(\zSigma+\mu\idem)\!\zcdot\!\xg\zjm\ge0,
\label{eq:delta2}
\end{equation}
with
\begin{equation}
\zD=\half(\xg\zv+(\xg\zv)^{\trans})
\end{equation}
the symmetric \emph{strain-rate}.

%%%%%%%%%%%%%%%%%%%%%%%%%%%%%%%%%%%%%%%%%%%%%%%%%%%%%%%%%%%%
\subsection{Constitutive equations}

We consider the cases of compressible and incompressible mixtures separately.

%%%%%%%%%%%%%%%%%%%%%%%%%%%%%%%
\subsubsection{Compressible mixture}

To the lists of dependent and independent constitutive variables we add
the convective stress $\zT$ and the mixture density $\varrho$ and
strain-rate $\zD$, respectively, so that
\begin{equation}
(\psi,\mskip2mu\mu,\mskip2mu\zSigma,\mskip2mu\zh,\mskip2mu\zT)
=\bigl(\hat{\psi}(z),\mskip2mu\hat{\mu}(z),\mskip2mu\hat{\zSigma}(z),
\mskip2mu\hat{\zh}(z),\mskip2mu\hat{\zT}(z)\bigr),
%\left.
%\begin{split}
%\psi&=\hat{\psi}(\varrho,\conc,\zjm,\zD),
%\qquad \zSigma=\hat{\zSigma}(\varrho,\conc,\zjm,\zD),
%\qquad\zT=\hat{\zT}(\varrho,\conc,\zjm,\zD).\\
%\mu&=\hat{\mu}(\varrho,\conc,\zjm,\zD),
% \qquad  \mskip9mu
%\zh=\hat{\zh}(\varrho,\conc,\zjm,\zD),
%\end{split} \quad
%\right\}
\label{eq:cr2}
\end{equation}
where, for brevity, we have introduced
\begin{equation}
z=(\varrho, \conc, \zjm, \zD).
\end{equation}

Inserting the constitutive relations (\ref{eq:cr2}) into the local energy
imbalance (\ref{eq:delta2}), we arrive at the functional inequality
\begin{multline}
\varrho\bigl(\frac{\partial\hat{\psi}}{\partial\conc}(z)
-\hat{\mu}(z)\bigr)\dot{\conc}
+\varrho\frac{\partial\hat{\psi}}{\partial\zjm}(z)\!\zcdot\!\skew3\dot{\zjm}
%\\[4pt]
+\varrho\frac{\partial\hat{\psi}}{\partial\zD}(z)
\!\zcdot\!\skew3\dot{\zD}
%%+\frac{\partial\hat{\psi}}{\partial\xg\conc}(z)\!\zcdot\!
%%\zdot{\overline{\xg \conc}}
+\hat{\zh}(\varrho,\conc,\zjm,\zD)\!\zcdot\!\zjm
\\[4pt]-\bigl(\hat{\zT}(z)
-\varrho^2\frac{\partial\hat{\psi}}{\partial\varrho}
(z)\idem\bigr)\!\cdot\!\zD
%\\[4pt]
-\bigl(\hat{\zSigma}(z)
+\hat{\mu}(z)\idem\bigr)
\!\zcdot\!\xg\zjm
 \le  0.
\label{eq:dissipation_imbalance_with_constitutive_relations2}
\end{multline}

Hence, proceding as in the our treatment of the theory without mixture
convection, we find that:
\begin{itemize}
\item[({\em i\/})] the constitutive response functions $\hat{\psi}$,
$\hat{\mu}$, and $\hat{\zSigma}$ delivering the mixture energy density
$\psi$, the mixture diffusion potential $\mu$, and the solute microstress
$\zSigma$ must be independent of $\zjm$, with
\begin{equation}
\psi=\hat{\psi} (\varrho,\conc),\qquad
\mu=\frac{ \partial}{\partial \conc} \hat{\psi} (\varrho,\conc),
\qquad\text{and}\qquad
\zSigma=-\hat{\mu}(\varrho,\conc) \idem;
\label{eq:basic_model_restrictions2}
\end{equation}
\item[({\em ii\/})] the constitutive response function $\hat{\zh}$ for the
internal solute body force density $\zh$ and the mixture stress $\zT$ must be
consistent with the \emph{residual inequality}
\begin{equation}
\hat{\zh}(\varrho,\conc,\zjm,\zD)\!\zcdot\!\zjm
 -{\zT}_\vis(\varrho,\conc,\zjm,\zD) \!\zcdot\!\zD \le0,
\label{eq:residual_inequality2}
\end{equation}
in which
\begin{equation}
\hat{\zT}_\vis(\varrho,\conc,\zjm,\zD)=\varrho^2\frac{\partial}{\partial\varrho}
\hat{\psi}(\varrho,\conc)\idem+\zT
\end{equation}
determines the dissipative contribution ${\zT}_\vis$ to the mixture
stress.
\end{itemize}

Further, granted smoothness of the response functions $\hat{\zh}$ and
$\hat{\zT}_\vis$, the previously employed result of Gurtin \& Voorhees
\cite{gurtin93} yields a general solution
\begin{equation}
\left\delimiter0
\begin{split}
\zh&=-\zZ(\varrho,\conc,\zjm,\zD)\mskip1mu\zjm
     -\zY(\varrho,\conc,\zjm,\zD)\mskip1mu\zD,
%\zh=&-\boldsymbol{\sf Z}(\varrho,\conc,\zjm,\zD)\,\zjm -
%\boldsymbol{\sf Y}(\varrho,\conc,\zjm,\zD)\,\zD
\\[4pt]
\zT_\vis&=\phantom{-}
   \zU(\varrho,\conc,\zjm,\zD)\mskip1mu\zjm
  +\zV(\varrho,\conc,\zjm,\zD)\mskip1mu\zD,
%\zT_\vis=&\boldsymbol{\sf U}(\varrho,\conc,\zjm,\zD)\,\zjm+
%\boldsymbol{\sf V}(\varrho,\conc,\zjm,\zD)\,\zD
\label{eq:solution_of_residual_inequality2}
\end{split}
\right\}
\end{equation}
of (\ref{eq:residual_inequality2}), where $\zZ$, $\zY$, $\zU$, and $\zV$
are isotropic due to objectivity and must
obey
\begin{equation}
\zjm\!\cdot\! \zZ(\varrho,\conc,\zjm,\zD)\mskip1mu\zjm
+\zjm\!\cdot\! \zY(\varrho,\conc,\zjm,\zD)\mskip1mu\zD
+\zD\!\cdot\! \zU(\varrho,\conc,\zjm,\zD)\mskip1mu\zjm
+\zD\!\cdot\! \zV(\varrho,\conc,\zjm,\zD)\mskip1mu\zD \ge0
\end{equation}
for all $(\varrho,\conc,\zjm,\zD)$. Despite their objectivity, these
coefficients need not reduce to scalars.

In this case, the full system of governing equations is
\begin{equation}
\left\delimiter0
\begin{split}
\dot{\varrho}&=-\varrho\mskip1mu\xd\zv,
\phantom{\frac{\partial}{\partial}}\\
\varrho\mskip1mu\dot{\conc}&=-\xd\zjm+m,
\phantom{\frac{\partial}{\partial}}\\
    \varrho\mskip1mu\skew2\dot{\zv}&=
   -\xg\bigl(\varrho^2\frac{\partial\hat{\psi}}{\partial\varrho}
     (\varrho,\conc)\bigr)%\\
%&\hskip1in
  +\xd\bigl(\zU(\varrho,\conc,\zjm,\zD)\zjm+\zV(\varrho,\conc,\zjm,\zD)\zD
\bigr) +\zb,\\ {\zZ}(\varrho,\conc,\zjm,\zD)\mskip1mu\zjm &=
   -\xg \bigl(\frac{\partial\hat{\psi}}{\partial\conc}
(\varrho,\conc)\bigr) -\zY(\varrho,\conc,\zjm,\zD) \zD     +\zell,
\label{eq:}
\end{split}
\right\}
\end{equation}

These equations are susbstantially more complicated than those for the case of
simple diffusion without convection. However,
if we require that $\hat{\zh}$ and $\hat{\zT}_\vis$ be linear functions
of $(\zjm,\zD)$, the general result
(\ref{eq:solution_of_residual_inequality2}) specializes to yield
\begin{equation}
\left\delimiter0
\begin{split}
\zh&=-{\zeta}(\varrho,\conc)\mskip1mu\zjm,
\\[4pt]
\zT_\vis&=\phantom{}2{\eta}_1(\varrho,\conc)\mskip1mu\zD
+{\eta}_2(\varrho,\conc)\mskip1mu(\tr\zD)\idem,
\label{eq:solution_of_residual_inequality3}
\end{split}
\right\}
\end{equation}
with
\begin{equation}
\zeta(\varrho,\conc)\ge0,
\qquad
\eta_1(\varrho,\conc)\ge0,\qquad \text{and}\qquad
 2\eta_1(\varrho,\conc)+3{\eta}_2(\varrho,\conc)\ge0
\end{equation}
for all $(\varrho,\conc)$. In this case, the full system of governing
equations has the form
\begin{equation}\left\delimiter0\begin{split}
\dot{\varrho}&=-\varrho\,\xd\zv,
\phantom{\frac{\partial}{\partial}}\\
\varrho\mskip1mu\dot{\conc}&=-\xd\zjm+m,
\phantom{\frac{\partial}{\partial}}\\
    \varrho\mskip1mu\skew2\dot{\zv}&=
   -\xg\bigl(\varrho^2\frac{\partial\hat{\psi}}{\partial\varrho}
     (\varrho,\conc)\bigr)
    +\xd\bigl(2{\eta_1}(\varrho,\conc)\zD\bigr)
+\xg\bigl(\eta_2(\varrho,\conc)\xd\zv\bigr) +\zb, \\
%&\hskip1.437in   +\xg\bigl(\eta_2(\varrho,\conc)\xd\zv\bigr) +\zb,\\
{\zeta}(\varrho,\conc)\mskip1mu\zjm &=
   -\xg \bigl(\frac{\partial\hat{\psi}}{\partial\conc}(\varrho,\conc)\bigr)
     +\zell,
\label{eq:compressible_model_final_equations}\end{split}\ \right\}
\end{equation}
which constitutes a system of coupled equations for the fields
$(\varrho,\conc,\zv,\zjm)$.

%%%%%%%%%%%%%%%%%%%%%%%%%%%%%%%
\subsubsection{Incompressible mixture}

If the convective mixture velocity obeys
\begin{equation}
\xd\zv=\tr\zD=0,
\label{eq:div_v=0}
\end{equation}
so that the motion of the mixture is \emph{isochoric}, we may without loss
of generality assume that $\varrho=1$ and consequently delete the mixture
density from the list of independent constitutive variables. For
convenience, we introduce the fields
\begin{center}
\begin{tabular}{cl}
$p$ & \emph{mixture pressure}, \\ [0.5ex]
$\zS$   & \emph{mixture extra stress},        \\ [0.5ex]
\end{tabular}
\end{center}
defined by the decomposition
\begin{equation}
\zT=-p\idem+\zS,
\qquad
p=-\textstyle{\frac{1}{3}}\tr\zT,
\qquad
\tr\zS=0,
\end{equation}
of the mixture stress into a constitutively indeterminate \emph{pressure}
$p$ that reacts to the constraint (\ref{eq:div_v=0}) and a traceless but
constitutively determinate \emph{extra stress} $\zS$.

We then assume that
\begin{equation}
(\psi,\mskip2mu\mu,\mskip2mu\zSigma,\mskip2mu\zh,\mskip2mu\zS)
=\bigl(\hat{\psi}(z),\mskip2mu\hat{\mu}(z),\mskip2mu\hat{\zSigma}(z),
\mskip2mu\hat{\zh}(z),\mskip2mu\hat{\zS}(z)\bigr),
%\left.
%\begin{split}
%\psi&=\hat{\psi}(\conc,\zjm,\zD),
%\qquad
%\zSigma=\hat{\zSigma}(\conc,\zjm,\zD),
%\qquad
%\zS=\hat{\zS}(\conc,\zjm,\zD).
%\\
%\mu&=\hat{\mu}(\conc,\zjm,\zD),
% \qquad\mskip9mu
%\zh=\hat{\zh}(\conc,\zjm,\zD),
%\end{split}
%\right\}
\label{eq:cr3}
\end{equation}
with
\begin{equation}
z=(\conc,\zjm,\zD),
\end{equation}
and, requiring that these relations satisfy the energy imbalance
(\ref{eq:delta2}) in all processes, find that the constitutive response
functions $\hat{\psi}$, $\hat{\mu}$, and $\hat{\zSigma}$ delivering the
mixture energy density $\psi$, the solute diffusion potential $\mu$, and the
solute microstress $\zSigma$ must, as in
(\ref{eq:basic_model_restrictions}), be independent of the diffusive mass
flux $\zjm$ and the strain-rate $\zD$, so that
\begin{equation}
\psi=\hat{\psi}(c),
\qquad
\mu=\hat{\psi}{}^{\prime}(\conc),
\qquad\text{and}\qquad
\zSigma=-\hat{\psi}{}^{\prime}(\conc)\idem.
\end{equation}
Further, granted that the constitutive response functions $\hat{\zh}$ and
$\hat{\zS}$ for the internal solute body microforce $\zh$ and the traceless
component $\zS$ of the mixture stress are smooth functions of
$(\conc,\zjm,\zD)$, we find that
\begin{equation}
\left\delimiter0
\begin{split}
\zh&=-\zZ(\conc,\zjm,\zD)\mskip1mu\zjm-\zY(\conc,\zjm,\zD)\mskip1mu\zD,
\\[4pt]
\zS&=\phantom{-}
\zU(\conc,\zjm,\zD)\mskip1mu\zjm+\zV(\conc,\zjm,\zD)\mskip1mu\zD,
\label{eq:solution_of_residual_inequality4}
\end{split}
\right\}
\end{equation}
where $\zZ$, $\zY$, $\zU$, and $\zV$ must obey
\begin{equation}
\zjm\!\cdot\!\zZ(\conc,\zjm,\zD)\mskip1mu\zjm
+\zjm\!\cdot\!\zY(\conc,\zjm,\zD)\mskip1mu\zD
+\zD\!\cdot\!\zU(\conc,\zjm,\zD)\mskip1mu\zjm
+\zD\!\cdot\!\zV(\conc,\zjm,\zD)\mskip1mu\zD\ge0
\end{equation}
for all $(\conc,\zjm,\zD)$.

If, as in the compressible case considered above, we require that
$\hat{\zh}$ and $\hat{\zS}$ be linear functions of $(\zjm,\zD)$, the general
result (\ref{eq:solution_of_residual_inequality4}) reduces to
\begin{equation}
\left\delimiter0
\begin{split}
\zh&=-{\zeta}(\conc)\mskip1mu\zjm,
\\[4pt]
\zS&=\mskip3mu\phantom{}2\mskip1mu{\eta}_1(\conc)\mskip1mu\zD,
\label{eq:solution_of_residual_inequality5}
\end{split}
\right\}
\end{equation}
with
\begin{equation}
\zeta(\conc)\ge0
\qquad\text{and}\qquad
\eta_1(\conc)\ge0
\end{equation}
for all $\conc$. In this case, the full system of governing equations is:
\begin{equation}\left\delimiter0\begin{split}
0 &=\xd\zv, \\
\dot{\conc}&=-\xd\zjm+m,\\
    \skew2\dot{\zv}&= -\xg p
    +\xd\bigl(2{\eta}_1(\conc)\zD\bigr) +\zb,\\
{\zeta}(\conc)\mskip1mu\zjm &=
    -\xg \bigl( \hat{\psi}^\prime(\conc)\bigr)
     +\zell,
\label{eq:incompressible_model_final_equations}\end{split}\ \right\}
\end{equation}
which constitutes a system of coupled equations for the fields
$(p,\conc,\zv,\zjm)$.

%%%%%%%%%%%%%%%%%%%%%%%%%%%%%%%%%%%%%%%%%
\section{Discussion}
\label{sect:discussion}

Our theory is based upon the interpretation of the diffusive  flux as a
generalized velocity and the introduction of corresponding
power-conjugate forces, which we call microforces, into an otherwise
conventional continuum-mechanical description. A systematic,
thermodynamically consistent derivation  generates a final system of
governing equations---(\ref{eq:basic_model_final_equations}) when
convection is ignored, and
(\ref{eq:compressible_model_final_equations}) or
(\ref{eq:incompressible_model_final_equations}) when convection is
taken into account---that constitutes a generalization of common
convection-diffusion equations, a generalization that accounts for
constitutive nonlinearities and external forces acting on the diffusing
constituents.

In the absence of mixture convection, we find that the constitutive
response of the material is determined by the provision of two scalar
functions, one for the energy density and the other for the frictional
resistance. This result generalizes naturally to account for dissipation
associated with the convective motion of the mixture. Importantly, we
arrive at our theory without introducing a constitutive relation such as
(\ref{eq:constitutive}) for the diffusive flux.

Additionally, our proposed format also leads to a generalization,
(\ref{eq:generalized_stokes_einstein}), of the classical
relation between diffusion coefficient and frictional resistance. We
find that the classical form for the  relation
(\ref{eq:diffusion_coef}) holds  when the free-energy density is given
by the special constitutive relation (\ref{eq:assumption1b}).

A distinction between the final governing equations of our theory and
those arising from more standard derivations of diffusion equations is
that, in our approach, a vector-valued microforce balance provides a
generalization between diffusive solute flux and gradient of the solute
diffusion potential. Furthermore, consistent with Nernst's treatment of
osmotic forces, the gradient of solute diffusion potential emerges
naturally as the diffusive force in this balance. This result for the
diffusive driving force is consistent with Batchelor's \cite{Batchelor76}
``thermodynamic force'' on particles in suspension. Typically, however,
this diffusive force is not formally included in force balances, but,
rather, a diffusive flux is added \emph{ad hoc} to the total flux. We
point out that Bird, Armstrong \& Hassager \cite{Bird77} include such a
diffusive force in their force balance for polymer suspensions.

%%%%%%%%%%%%%%%%%%%%%%%%%%%%%%%%%%%%%%%%%%%%%%%%%%%%%%%%%%%%%%%%%%%%%
\section*{Acknowledgements}

This work was partially supported by the U.\ S.\ Department of Energy and
the EPSRC (UK).

%%%%%%%%%%%%%%%%%%%%%%%%%%%%%%%%%%%%%%%%%%%%%%%%%


\begin{thebibliography}{99}

{\renewcommand{\baselinestretch}{0.95}
\footnotesize

\vspace{-4pt}\bibitem{Landau59}
L.\ D.\ Landau \& E.\ M.\ Lifshitz, \emph{Fluid Mechanics}, Pergamon
Press, London, 1959.

\vspace{-4pt}\bibitem{Degroot83} S.\ R.\ de Groot \& P.\ Mazur,
\emph{Non-Equilibrium Thermodynamics}, Dover, New York, 1983.

\vspace{-4pt}\bibitem{Cussler84}
E.\ L.\ Cussler, \emph{Diffusion, Mass Transfer in Fluid Systems},
Cambridge University Press, Cambridge, 1984.

\vspace{-4pt}\bibitem{Fick55}
A.\ Fick, \"Uber Diffusion, \emph{Poggendorffs
Annalen} {\bf 94} (1855), 59--86.

\vspace{-4pt}\bibitem{Germain}
P.\ Germain, Sur l'application de la m\'ethode des puissances virtuelles
en m\'ecanique des milieux continus, \emph{Comptes Rendus de
l'Acad\'emie des Sciences, s\'erie A} {\bf 274} (1973), 1051--1055.

\vspace{-4pt}\bibitem{Antman}
S.\ S.\ Antman \& J.\ E.\ Osborn, The principle of virtual work and
integral laws of motion, \emph{Archive for Rational Mechanics and
Analysis} {\bf 69} (1979), 231--262.

\vspace{-4pt}\bibitem{Maugin}
G.\ A.\ Maugin, The principle of virtual power in continuum
mechanics---application to coupled fields, \emph{Acta Mechanica} {\bf 35}
(1980), 1--70.

\vspace{-4pt}\bibitem{Fremond}
M.\ Fr\'emond, Endommagement et principe des puissances virtuelles, {\em
Comptes Rendus de l'Acad\'emie des Sciences, s\'erie II\/} {\bf 317}
(1993), 857--863.

\vspace{-4pt}\bibitem{E61}
J.\ L.\ Ericksen, Conservation laws for liquid crystals, {\em
Transactions of the Society of Rheology\/} {\bf 5} (1961), 23--34.

\vspace{-4pt}\bibitem{GC72}
M.\ A.\ Goodman \& S.\ C.\ Cowin, A continuum theory for granular
materials, {\em Archive for Rational Mechanics and Analysis\/} {\bf 44}
(1972), 249--266.

\vspace{-4pt}\bibitem{CP-G}
G.\ Capriz \& P.\ Podio-Guidugli, Structured continua from a Lagrangian
point of view, {\em Annali di Matematica Pura ed Applicata\/} {\bf 135}
(1983), 1--25.

\vspace{-4pt}\bibitem{C89}
G.\ Capriz, {\em Continua with Microstructure}, Springer-Verlag, New
York, 1989.

\vspace{-4pt}\bibitem{FG}
E.\ Fried \& M.\ E.\ Gurtin, Dynamic solid-solid transitions with phase
characterized by an order parameter, \emph{Physica D} {\bf 68} (1993),
326--343.

\vspace{-4pt}\bibitem{F} E.\ Fried, Continua described by a microstructural
field, \emph{Zeitschrift f\"ur angewandte Mathematik und Physik} {\bf 47}
(1996), 168--175.

\vspace{-4pt}\bibitem{G}
M.\ E.\ Gurtin, Generalized Ginzburg-Landau and Cahn-Hilliard equations based
on a microforce balance, \emph{Physica D} {\bf 92} (1996), 178--192.

\vspace{-4pt}\bibitem{Nernst88}
W.\ Nernst, Zur Kinetik der in L\"osung befindlichen K\"orper: I. Theorie
der Diffusion, \emph{Zeitschrift f\"ur physikalische Chemie} {\bf 2}
(1888), 613--637.

\vspace{-4pt}\bibitem{Hoff87}
J.\ H.\ van't Hoff, Die Rolle des osmotischen Druckes in der Analogie
zwischen L\"osungen und Gasen, \emph{Zeitschrift f\"ur physikalische
Chemie} {\bf 1} (1887), 481--508.

\vspace{-4pt}
\bibitem{Planck90}
M.\ Planck, Ueber die Erregung von Electricit\"at und W\"arme in
Electrolyten, \emph{Annalen der Physik und Chemie} (\emph{Neue Folge})
{\bf 39} (1890), 161--186.

\vspace{-4pt}\bibitem{einstein05}
A.\ Einstein, \"{U}ber die von der molekularkinetischen Theorie der
W\"{a}rme geforderte Bewegung von in ruhenden Fl\"{u}ssigkeiten
suspendierten Teilchen, \emph{Annalen der Physik} {\bf 17} (1905),
549--560.

\vspace{-4pt}\bibitem{einstein06}
A.\ Einstein, Zur Theorie der Brownschen Bewegung, \emph{Annalen der
Physik} {\bf 19} (1906), 371--381.

\vspace{-4pt}\bibitem{einstein08}
A.\ Einstein, Elementare Theorie der Brownschen Bewegung,
\emph{Zeitschrift f\"ur Elektrochemie} {\bf 14} (1908), 235--239.

\vspace{-4pt}\bibitem{einstein06b}
A.\ Einstein, Eine neue Bestimmung der Molek\"uldimensionen,
\emph{Annalen der Physik} {\bf 19} (1906), 289--306.

\vspace{-4pt}\bibitem{Debye62}
A.\ Prock \& G.\ McConkey, \emph{Topics in Chemical Physics: Based
on the Harvard Lectures of Peter J.\ W.\ Debye},
Elsevier, Amsterdam, 1962.

\vspace{-4pt}\bibitem{smoluchowski15}
M.\ von Smoluchowski, \"Uber {B}rownsche {M}olekularbewegung unter
{E}inwirkung \"au{\ss}erer {K}r\"afte und deren {Z}usammenhang mit der
verallgemeinerten {D}iffusionsgleichung, \emph{Annalen der Physik} {\bf
48} (1915), 1103--1112.

\vspace{-4pt}\bibitem{boltzmann72}
L.\ Boltzmann, Weitere {S}tudien \"uber das {W}\"armegleichgewicht unter
Gasmolek\"ulen. \emph{Sitzungsberichte der Akademie der Wissenschaften,
Wien} {\bf 66} (1872), 275--370.

\vspace{-4pt}\bibitem{debye13}
P.\ Debije, Zur Theorie der anomalen Dispersion im Gebiete der langwelligen
elektrischen Strahlung, \emph{Berichte der deutschen physikalischen
Gesellschaft} {\bf 15} (1913), 777--793.

\vspace{-4pt}\bibitem{debye29}
P.\ Debye, \emph{Polar Molecules}, Chemical Catalog Company, New York,
1929.

\vspace{-4pt}
\bibitem{Giesekus58}
H.\ Giesekus, Die rheologische Zustandsgleichung, \emph{Rheologica Acta}
{\bf 1} (1958), 2--20.

\vspace{-4pt}
\bibitem{Doi86}
M.\ Doi \& S.\ F.\ Edwards, \emph{The Theory of Polymer Dynamics},
Clarendon Press, Oxford, 1986.

\vspace{-4pt}
\bibitem{Bird77}
R.\ B.\ Bird, R.\ C.\ Armstrong \& O.\ Hassager, \emph{The Dynamics of
Polymeric Liquids}, Volume 2, Wiley, New York, 1977.

\vspace{-4pt}\bibitem{CN}
B.\ D.\ Coleman \& W.\ Noll, The thermodynamics of elastic materials with heat
conduction and viscosity, \emph{Archive for Rational Mechanics and Analysis}
{\bf 13} (1963), 167--178.

\vspace{-4pt}\bibitem{gurtin93}
M.\ E.\ Gurtin \& P.\ W.\ Voorhees, The continuum mechanics of coherent
two-phase elastic solids with mass transport, \emph{Proceedings of the
Royal Society of London A} {\bf 444} (1993), 323--343.

\vspace{-4pt}\bibitem{Stokes}
G.\ G.\ Stokes, On the effect of internal friction of fluids on the motion
of pendulums, \emph{Transactions of the Cambridge Philosophical Society}
{\bf 9} (1851), 8--106 = \emph{Mathematical and Physical Papers},
Volume 3, Cambridge University Press, Cambridge, 1901.

\vspace{-4pt}
\bibitem{Batchelor76}
G.\ K.\ Batchelor, Brownian diffusion of particles with hydrodynamic
interaction, \emph{Journal of Fluid Mechanics} {\bf 74} (1976), 1--29.

\par}

\end{thebibliography}
\end{document}